\documentclass[ALICE,manyauthors]{cernphprep}

\usepackage[comma,square,numbers,sort&compress]{natbib}
\usepackage{hyperref}
\usepackage{lineno}
\usepackage{xspace}
\usepackage{xcolor}
\usepackage{textcomp}
\usepackage[T1]{fontenc}

\newcommand{\snn}{\ensuremath{\sqrt{s_{\mathrm{NN}}}}\xspace}
\newcommand{\seven}{\ensuremath{\sqrt{s}=7} TeV\xspace}

\newcommand{\hyp}    {\ensuremath{^{3}_{\Lambda}\mathrm H}\xspace}

\newcommand{\PbPb}{Pb--Pb\xspace}

\newcommand{\permille}{\textperthousand\xspace}

\newcommand{\pt}   {\ensuremath{p_{\mathrm{T}}}\xspace}
\newcommand{\mt}   {\ensuremath{m_{\mathrm{T}}}\xspace}

\newcommand{\dd}     {\ensuremath{\mathrm{d}}\xspace}
\newcommand{\dndy} {\mbox{\ensuremath{\dd N/\dd y}}\xspace}
\newcommand{\dndeta} {\mbox{\ensuremath{\dd N_{ch}/\dd \eta}}\xspace}
\newcommand{\meanpt}     {\ensuremath{\langle p_{\mathrm{T}}\rangle}\xspace}

\newcommand{\dedx}{\mbox{\ensuremath{\dd E/\dd x}}\xspace}
\newcommand{\dvdy}{\mbox{\ensuremath{\dd V/\dd y}}\xspace}

\begin{document}%

\begin{titlepage}
\PHyear{2020}
\PHnumber{025}      
\PHdate{03 March}  
%

\title{(Anti-)Deuteron production in pp collisions at $\sqrt{s}=13$ TeV}
\ShortTitle{(Anti-)Deuteron production in pp collisions at $\sqrt{s}=13$ TeV}   

\Collaboration{ALICE Collaboration\thanks{See Appendix~\ref{app:collab} for the list of collaboration members}}
\ShortAuthor{ALICE Collaboration} 

\begin{abstract}
The study of (anti-)deuteron production in pp collisions has proven to be a powerful tool to investigate the formation mechanism of loosely bound states in high energy hadronic collisions. In this paper the production of $\text{(anti-)deuterons}$ is studied as a function of the charged particle multiplicity in inelastic pp collisions at $\sqrt{s}=13$ TeV using the ALICE experiment. Thanks to the large number of accumulated minimum bias events, it has been possible to measure (anti-)deuteron production in pp collisions up to the same charged particle multiplicity ($\dndeta\sim26$) as measured in p--Pb collisions at similar centre-of-mass energies. Within the uncertainties, the deuteron yield in pp collisions resembles the one in p--Pb interactions, suggesting a common formation mechanism behind the production of light nuclei in hadronic interactions.
In this context the measurements are compared with the expectations of coalescence and Statistical Hadronisation Models (SHM).
\end{abstract}
\end{titlepage}
\setcounter{page}{2}

\section{Introduction}

High energy collisions at the Large Hadron Collider (LHC) create a suitable environment for the production of light (anti-)nuclei.
In ultra-relativistic heavy-ion collisions light (anti-)nuclei are abundantly produced~\cite{Adam:2015vda,Adler:2001uy,Adler:2004uy}, 
but in elementary pp collisions their production is lower \cite{Alper:1973my,Henning:1977mt,Adam:2015vda,Acharya:2017fvb}. As a consequence, there are only few detailed measurements of (anti-)nuclei production rate in pp collisions.
However, with the recently collected large data sample it is now possible to perform more differential measurements of light (anti-)nuclei production as a function of multiplicity and transverse momentum. In this paper, we present the detailed study of the multiplicity dependence of (anti-)deuteron production in pp collisions at  $\sqrt{s} =$ 13 TeV, the highest collision energy so far delivered at the LHC.

The production mechanism of light (anti-)nuclei in high energy hadronic collisions is not completely understood. However, two groups of models have turned out to be particularly useful, namely Statistical Hadronisation Models (SHM) and coalescence models. The SHMs, which assume particle production according to the thermal equilibrium expectation, have been very successful in explaining the yields of light (anti-)nuclei along with other hadrons in Pb--Pb collisions~\cite{Acharya:2017bso}, suggesting a common chemical freeze-out temperature for light (anti-)nuclei and other hadron species. 
The ratio between the \pt-integrated yields of deuterons and protons (d/p ratio) in Pb--Pb collisions remains constant as a function of centrality, but rises in pp and p--Pb collisions with increasing multiplicity, finally reaching the value observed in Pb--Pb \cite{Adam:2015vda,Acharya:2019rgc,Acharya:2019rys}.  
The constant d/p ratio in Pb--Pb collisions as a function of centrality is consistent with thermal production, suggesting that the chemical freeze-out temperature in Pb--Pb collisions does not vary with centrality~\cite{Sharma:2018jqf}. 
Assuming thermal production in pp collisions as well, the lower d/p ratio would indicate a lower freeze-out temperature \cite{Sharma:2018jqf}. On the other hand, the ratio between the \pt-integrated yields of protons and pions (p/$\pi$ ratio) does not show a significant difference between pp and Pb--Pb collisions \cite{Abelev:2013vea,Adam:2015qaa}. 
Also, for p--Pb collisions the freeze-out temperature obtained with SHMs using only light-flavoured particles is constant with multiplicity and its value is similar to that obtained in Pb--Pb collisions \cite{Sharma:2018owb}. 
Thus, the increase of the d/p ratio with multiplicity for smaller systems cannot be explained within the scope of the grand-canonical SHM as is done in case of Pb--Pb.
It is also not consistent with a simple SHM that 
the d/p and p/$\pi$ ratios 
behave differently as a function of multiplicity even though numerator and denominator differ in both cases by one unit of baryon number. 
Nonetheless, a process similar to the canonical suppression of strange particles might be worth considering also for baryons. A recent calculation within the SHM approach with exact conservation of baryon number, electric charge, and strangeness 
focuses on this aspect~\cite{Vovchenko:2018fiy}.

In coalescence models (anti-)nuclei are formed by nucleons close in phase-space~\cite{Kapusta:1980zz}.
In this approach, the coalescence parameter $B_2$ quantitatively describes the production of $\text{(anti-)deuterons}$. $B_2$ is defined as
\begin{equation}
    B_{2}\left(p_{\mathrm{T}}^p\right) = E_{d}\frac{\mathrm{d}^3N_{d}}{\mathrm{d}p_{d}^3}\bigg/\left(E_{p}\frac{\mathrm{d}^3N_{p}}{\mathrm{d}p_{p}^3}\right)^2 = \frac{1}{2\pi p_{\mathrm{T}}^d}\frac{\mathrm{d}^2N_{d}}{\mathrm{d}y\mathrm{d}p^{d}_{\mathrm{T}}} \; \bigg/ \left(\frac{1}{2\pi p_{\mathrm{T}}^p}\frac{\mathrm{d}^2N_{p}}{\mathrm{d}y\mathrm{d}\pt^{p}}\right)^2 ,
\end{equation}
where $E$ is the energy, $p$ is the momentum, \pt~is the transverse momentum and $y$ is the rapidity. The labels $p$ and $d$ are used to denote properties related to protons and deuterons, respectively. The invariant spectra of the $\text{(anti-)protons}$ are evaluated at half of the transverse momentum of the deuterons, so that $p_{\mathrm{T}}^p = p_{\mathrm{T}}^d / 2$. Neutron spectra are assumed to be equivalent to proton spectra, since neutrons and protons belong to the same isospin doublet.
Since the coalescence process is expected to occur at the late stage of the collision, the parameter $B_2$ is related to the emission volume. In a simple coalescence approach, which describes the uncorrelated particle emission from a point-like source, $B_2$ is expected to be independent of \pt and multiplicity. However, it has been observed that $B_2$ at a given transverse momentum decreases as a function of multiplicity, suggesting that the nuclear emission volume increases with multiplicity \cite{Adler:2001uy,Anticic:2004yj,Acharya:2019rys}. 
In Pb--Pb collisions the $B_2$ parameter as a function of $p_T$ shows an increasing trend,
which is usually attributed to the position-momentum correlations caused by radial flow or hard scatterings \cite{Polleri:1997bp,Sharma:2018dyb}. Such an increase of $B_2$ as a function of \pt has in fact also been observed in pp collisions at \seven~\cite{Acharya:2017fvb}. 
However, if pp collisions are studied in separate intervals of multiplicity, $B_2$ is found to be almost constant as a function of \pt~\cite{Acharya:2019rgc}. Similarly, $B_2$ does not depend on \pt in multiplicity selected p--Pb collisions \cite{Acharya:2019rys}. 
Moreover, the highest multiplicities reached in pp collisions are comparable with those obtained in p--Pb collisions and not too far from peripheral Pb--Pb collisions. Therefore, the measure of $B_2$ as a function of \pt for finer multiplicity intervals in pp collisions at $\sqrt{s} =$~13~TeV gives the opportunity to compare different collision systems and to evaluate the dependence on the system size. 

The paper is organized as follows. Section~\ref{sec:detector} discusses the details of  the ALICE detector. Section~\ref{sec:eventselection} describes the data sample used for the analysis and the corresponding event and track selection criteria. Section~\ref{sec:analysis} presents the data analysis steps in detail, such as raw yield extraction and various corrections, as well as the systematic uncertainty estimation. In Section \ref{sec:results}, the results are presented and discussed. Finally, conclusions are given in Section~\ref{sec:conclusions}.

\section{The ALICE detector}
\label{sec:detector}
A detailed description of the ALICE detectors can be found in~\cite{Abelev:2014ffa} and references therein.
For the present analysis the main sub-detectors used are the V0, the Inner Tracking System (ITS), the Time Projection Chamber (TPC) and the Time-of-Flight (TOF), which are all located inside a 0.5~T solenoidal magnetic field. 

The V0 detector~\cite{Abbas:2013taa} is formed by two arrays of scintillation counters placed around the beampipe on either side of the interaction point: one covering the pseudorapidity range $2.8 < \eta < 5.1$~\mbox{(V0A)}  and the other one covering $-3.7 < \eta < -1.7$~\mbox{(V0C)}.
The collision multiplicity is estimated using the counts in the V0 detector, which is also used as trigger detector. More details will be given in Section \ref{sec:eventselection}.
 
The ITS \cite{Aamodt:2010aa}, designed to provide high resolution track points in the proximity of the interaction region, is composed of three subsystems of silicon detectors placed around the interaction region with a cylindrical symmetry. The Silicon Pixel Detector (SPD) is the subsystem closest to the beampipe and is made of two layers of pixel detectors.
The third and the fourth layers consist of Silicon Drift Detectors (SDD), while the outermost two layers are equipped with double-sided Silicon Strip Detectors (SSD). The inner radius of the SPD, 3.9~cm, is essentially given by the radius of the beam pipe, while the inner field cage of the TPC limits the radial span of the entire ITS to be 43~cm. The ITS covers the pseudorapidity range $|\eta |<0.9$ and it is hermetic in azimuth.

The same pseudorapidity range is covered by the TPC \cite{Alme:2010ke}, which is the main tracking detector, consisting of a hollow cylinder whose axis coincides with the nominal beam axis. The active volume, filled with a Ne/CO$_2$/N$_2$ gas mixture (Ar/CO$_2$/N$_2$ in 2016), at atmospheric pressure, has an inner radius of about 85~cm, an outer radius of about 250~cm, and an overall length along the beam direction of 500~cm.
The gas is ionised by charged particles traversing the detector and the ionisation electrons drift, under the influence of a constant electric field of $\sim $ 400~V/cm, towards the endplates, where their position and arrival time are measured.
The trajectory of a charged particle is estimated using up to 159 combined measurements (clusters) of drift times and radial positions of the ionisation electrons.
The charged-particle tracks are then formed by combining the hits in the ITS and the reconstructed clusters in the TPC. 
The TPC is used for particle identification by measuring the specific energy loss (\dedx) in the TPC gas.

The TOF system \cite{Akindinov:2013tea} covers the full azimuth for the pseudorapidity interval $|\eta|<0.9$. The detector is based on the Multi-gap Resistive Plate Chambers (MRPCs) technology and it is located, with a cylindrical symmetry, at an average distance of 380 cm from the beam axis.
The particle identification is based on the difference between the measured time-of-flight and its expected value, computed for each mass hypothesis from track momentum and length. The overall resolution on the time-of-flight of particles is about 80~ps.

A precise starting signal for the TOF system can be also provided by the T0 detector, consisting of two arrays of Cherenkov counters, T0A and T0C, which cover the pseudorapidity regions $4.61 < \eta < 4.92$ and $3.28 < \eta < 2.97$, respectively~\cite{Adam:2016ilk}. Alternatively, the start time can be provided by the TOF itself or the bunch-crossing time can be used, as described in \cite{Adam:2016ilk}.

\section{Data sample}
\label{sec:eventselection}
The data samples used in this work consist of approximately 950 million minimum bias pp events collected during the LHC proton runs in 2016 and 2017. The data were collected using a minimum-bias trigger requiring at least one hit in both the V0 detectors. Moreover, the timing information of the V0 scintillators is used for the offline rejection of events triggered by interactions of the beam with the residual gas in the LHC vacuum pipe. To ensure the best possible performance of the detector, events with more than one reconstructed primary interaction vertex (pile-up events) were rejected.

The production of primary $\text{(anti-)deuterons}$ is measured around mid-rapidity. In particular, the spectra are provided within a rapidity window of $|y| < 0.5$. To ensure that all tracks have the maximal length, only those in the pseudorapidity interval  $|\eta| < 0.8$ are selected. In order to guarantee good track momentum and \dedx resolution in the relevant \pt ranges, the selected tracks are required to have at least 70 reconstructed points in the TPC and two points in the ITS. In addition, at least one of the ITS points has to be measured by the SPD in order to assure for the selected tracks a resolution better than 300 $\mu$m on the distance of closest approach to the primary vertex in the plane perpendicular (DCA$_{xy}$) and parallel (DCA$_{z}$) to the beam axis~\cite{Abelev:2014ffa}. Furthermore, it is required that the $\chi^2$ per TPC reconstructed point is less than 4 and tracks originating from kink topologies of weak decays are rejected.

Data are divided into ten multiplicity classes, identified by a roman number from I to X, going from the highest to the lowest multiplicity. 
However, in this analysis classes IV and V are merged into a single class to achieve a better statistical precision.
The multiplicity classes are  determined from the sum of the V0 signal amplitudes and defined in terms of percentiles of the INEL$>0$ pp cross section, where INEL $>$ 0 events are defined as collisions with at least one charged particle in the pseudo-rapidity region $|\eta|<$1~\cite{Acharya:2019kyh}. The mean charged particle multiplicity $\left<\dndeta\right>$ for each class is reported in Table~\ref{tab:yields}. 

\section{Data analysis}
\label{sec:analysis}

\subsection{Raw yield extraction}
The identification of $\text{(anti-)deuterons}$ is performed with two different methods, depending on their transverse momentum. For $p_{\mathrm{T}} < $ 1 GeV/\textit{c}, the identification is done using a measurement of the \dedx in the TPC only. In particular, for each $p_{\mathrm{T}}$ interval the number of $\text{(anti-)deuterons}$ is extracted through a fit with a Gaussian with two exponential tails to the $n_{\sigma}$ distribution. Here, $n_{\sigma}$ is the difference between the measured TPC \dedx and the expected one for $\text{(anti-)deuterons}$ divided by the TPC \dedx resolution. However, for $p_{\mathrm{T}} \geq $ 1 GeV/\textit{c} it is more difficult to separate $\text{(anti-)deuterons}$ from other charged particles with this technique. Therefore, the particle identification in this kinematic region is performed using the TOF detector. The squared mass of the particle is computed as $m^{2} = p^{2}\left(t_{\mathrm{TOF}}^2/L^2 - 1/c^2\right)$, where $t_{TOF}$ is the measured time-of-flight, $L$ is the length of the track and $p$ is the momentum of the particle. In order to reduce the background, only the candidates with a d$E$/d$x$ measured in the TPC compatible within 3$\sigma$ with the expected value for a (anti-)deuteron are selected. The squared-mass-distributions are fitted with a Gaussian function with an exponential tail for the signal.
A significant background is present for $p_{\mathrm{T}} \geq $ 1.8 GeV/\textit{c} and is modelled with two exponential functions. In the range where the background is negligible, the raw yield is extracted by directly counting the candidates. Otherwise, the squared-mass distribution is fitted with the described model, using an extended-maximum-likelihood approach. The $\text{(anti-)deuteron}$ yield is then obtained by a fit parameter.
\subsection{Efficiency and acceptance correction}
\label{sec:efficiency}
A correction for the tracking efficiency and the detector acceptance must be applied to obtain the real yield. The correction is evaluated from Monte Carlo (MC) simulated events. The events are generated using the standard generator PYTHIA8 (Monash 2013)\cite{sjostrand2008brief}. However, PYTHIA8 does not handle the production of nuclei. Therefore, in each event it is necessary to inject $\text{(anti-)deuterons}$. In each pp collision one deuteron or one anti-deuteron is injected, randomly chosen from a flat rapidity distribution in the range $|y|<$ 1 and a flat $p_{\mathrm{T}}$ distribution in the range $p_{\mathrm{T}} \in [0,10] $ GeV/\textit{c}.  The correction is defined as the ratio between the number of reconstructed $\text{(anti-)deuterons}$ in the rapidity range $|y|<0.5$ and in the pseudorapidity interval $|\eta|<0.8$ and the number of generated ones in  $|y|<0.5$. The correction is computed separately for deuterons and anti-deuterons and for the TPC and TOF analyses.\\
Another correction is related to the trigger efficiency. All the selected events are required to have at least one  charged particle in the acceptance, i.e. in the pseudo-rapidity region $|\eta|<$1 (INEL $>$ 0)~\cite{Acharya:2019kyh}. Due to the imperfection of the trigger, some INEL $>$ 0 events are wrongly rejected (event loss). Consequently, all the $\text{(anti-)deuterons}$ produced in the erroneously rejected events are lost as well (signal loss). Therefore, it is necessary to correct the spectra for the event and the signal losses. Event loss is more relevant at low multiplicity and almost negligible at high multiplicity ($\sim$~12\% for multiplicity class X and $<$~1\permille for multiplicity class I). The corrections are computed from MC simulations, because both the number of rejected events and the number of $\text{(anti-)deuterons}$ produced in those same events are known. However, it is not possible to count the number of lost $\text{(anti-)deuterons}$ directly, because the artificial injection of one (anti-)deuteron per event will bias the number of lost candidates that can be extracted from this MC data set. Instead, the number of lost pions, kaons and protons are extracted from a different MC data set and then these values are extrapolated to the deuteron mass.
The standard transport code used in ALICE simulations is GEANT3. However, it is known from other ALICE analyses on nuclei that GEANT4 provides a more realistic transport of (anti-)nuclei. The GEANT3 response is hence scaled to the GEANT4 one to take into account this effect. Moreover, the spectra obtained with TOF are further corrected to take into account the TPC-TOF matching efficiency using a data-driven approach. This correction was evaluated for the analysis of the (anti-)deuteron production in the p--Pb data sample collected in 2013 \cite{Acharya:2019rys}. In that year not all the modules of the Transition Radiation Detector (TRD), which is located between the TPC and the TOF, were already installed. In this way it was possible to compute the effects of the presence of the TRD, comparing the (anti-)deuteron yields in the regions where the TRD modules were present and in those where they were not yet installed.
This correction was also verified with Run 2 data, by comparing the yields extracted with the TPC with those extracted with the TOF in the \pt region where both the techniques can be used.

\subsection{Subtraction of secondary deuterons}
Secondary deuterons are produced in the interaction of particles with the detector material and their contribution must be subtracted from the total measured deuteron yield. However, the production of secondary anti-deuterons is extremely rare due to baryon number conservation. Hence, the correction is applied only to the deuteron spectra.
The fraction of primary deuterons is evaluated via a fit to the DCA$_{xy}$ distribution of the data, as described in \cite{Adam:2015vda}. The template for primary deuterons is obtained from the measured DCA$_{xy}$ of anti-deuterons. The template from secondary deuterons is instead obtained from MC simulations. The production of secondary deuterons is more relevant at low $p_{\mathrm{T}}$ (at $p_{\mathrm{T}} = $ 0.7 GeV/\textit{c} the fraction of secondary deuterons is $\sim$ 40\%) and decreases exponentially with the transverse momentum ($<$ 5\% for $p_{\mathrm{T}} = $ 1.4 GeV/\textit{c}).
The only other possible contribution to secondary deuterons that is known is the decay $\hyp\rightarrow \mathrm{d} + \mathrm{p} +\pi$. However, \hyp production has not yet been observed in pp collisions and its production yield is therefore lower than that of $^3$He, which is less than a thousandth of the deuteron production rate \cite{Acharya:2017fvb}.
\subsection{Systematic uncertainties}
A list of all the sources of systematic uncertainty is shown in Table \ref{tab:systematics}. The values are reported for the multiplicity classes I and X, for the lowest and highest \pt values.\\
The track selection criteria are a source of systematic uncertainty. In this category we include all the contributions related to the single-track selection: DCA, number of clusters in the TPC and, for the TOF analysis, the width of the \dedx selection applied in the TPC.
These uncertainties are evaluated by varying the relevant selections, as done in \cite{Acharya:2019rgc}. At low $p_{\mathrm{T}}$ ($\pt < 1$ GeV/\textit{c}) the contribution is 2\% for deuterons due to the DCA$_z$ and DCA$_{xy}$ selections, which influence the estimation of the fraction of primary deuterons, while for anti-deuterons this systematic uncertainty is around 1\%. It increases with \pt and the growth is more pronounced for low multiplicity.
The systematic uncertainty on the signal extraction is evaluated by directly counting the (anti-)deuteron candidates. It is obtained by varying the interval in which the direct counting is performed. Its contribution is $\sim$ 1\% at low \pt and increases with \pt. Another source of systematic uncertainty is given by the incomplete knowledge of the material budget of the detector in the Monte Carlo simulations. The effect is evaluated by comparing different MC simulations in which the material budget was increased and decreased by 4.5\%. This value corresponds to the uncertainty on the determination of the material budget by measuring photon conversions. This particular systematic uncertainty is below 1\%. The imperfect knowledge of the hadronic interaction cross section of $\text{(anti-)deuterons}$ with the material contributes to the systematic uncertainty as well. Its effect is evaluated with the same data-driven approach used to investigate the  TOF-matching efficiency, as described in section \ref{sec:efficiency}. Half of the correction, corresponding to the 1$\sigma$ confidence interval,  is taken as its uncertainty contributing 4\% to the systematic uncertainty for deuterons and 7.5\% for anti-deuterons. Similarly, an uncertainty related to the ITS-TPC matching is considered. It is evaluated from the difference between the ITS-TPC matching efficiencies in data and MC and its contribution is less than 2.5\%.
Finally, a source of systematic uncertainties results from the signal loss correction. It is assumed to be half of the difference between the signal-loss correction (described in section \ref{sec:efficiency}) and 1. It is strongly dependent on the event multiplicity: it is negligible at high multiplicity (multiplicity classes from I to VII) and contributes up to 6\% in the lowest multiplicity class (class X). Where present, it decreases with \pt.
 \renewcommand\arraystretch{1.2}
 
 \begin{table}[ht]
   \centering
   \caption{Summary of the main contributions to the systematic uncertainties for the extreme multiplicity classes I and X. Values in brackets are referred to anti-deuterons. If they are not present, the systematic uncertainty is common for deuterons and anti-deuterons. More details about the sources of the uncertainties can be found in the text.}
   \vspace{10pt}
   \renewcommand{\arraystretch}{1.5}
    \begin{tabular}{clcclcc}
    \hline
    Source                               & \multicolumn{6}{c}{d ($\bar{\mathrm{d}}$)}\\
    \hline
    Multiplicity                         &  & \multicolumn{2}{c}{Class I}                                   &  & \multicolumn{2}{c}{Class X}\\
    \hline
    \pt (GeV/$c$)                        &  & 0.7                           & 3.8                           &  & 0.7          & 2.6         \\
    \hline
    Track selection                      &  & 2\% (1\%)                     & 2\% (3\%)                     &  & 2\% (1\%)    & 5\% (6\%)    \\
    Signal extraction                    &  & 1\%                           & 7\% (7\%)                     &  & 1\%          & 5\% (5\%)    \\
    Material budget                      &  &   $< 1\%$                           & $< 1\%$                     &  & $< 1\%$          & $< 1\%$          \\
    TPC-TOF matching                     &  & 4\% (7.5\%)                   & 4\% (7.5\%)                   &  & 4\% (7.5\%)  & 4\% (7.5\%)  \\
    ITS-TPC matching                     &  & 1\%                           & 2.5\%                         &  & 1\%          & 2.5\%        \\
    Signal Loss                          &  & -                             & -                             &  & 6\%          & 3\%          \\ \hline
    Total                                &  & 5\% (8\%)                     & 9\% (11\%)                    &  & 8\% (10\%)   & 10\% (12\%)  \\ \hline
    \end{tabular}
    \label{tab:systematics}
 \end{table}

\begin{figure}[ht]
	\centering
	\begin{subfigure}{}
	\centering
        \includegraphics[width=0.7\textwidth]{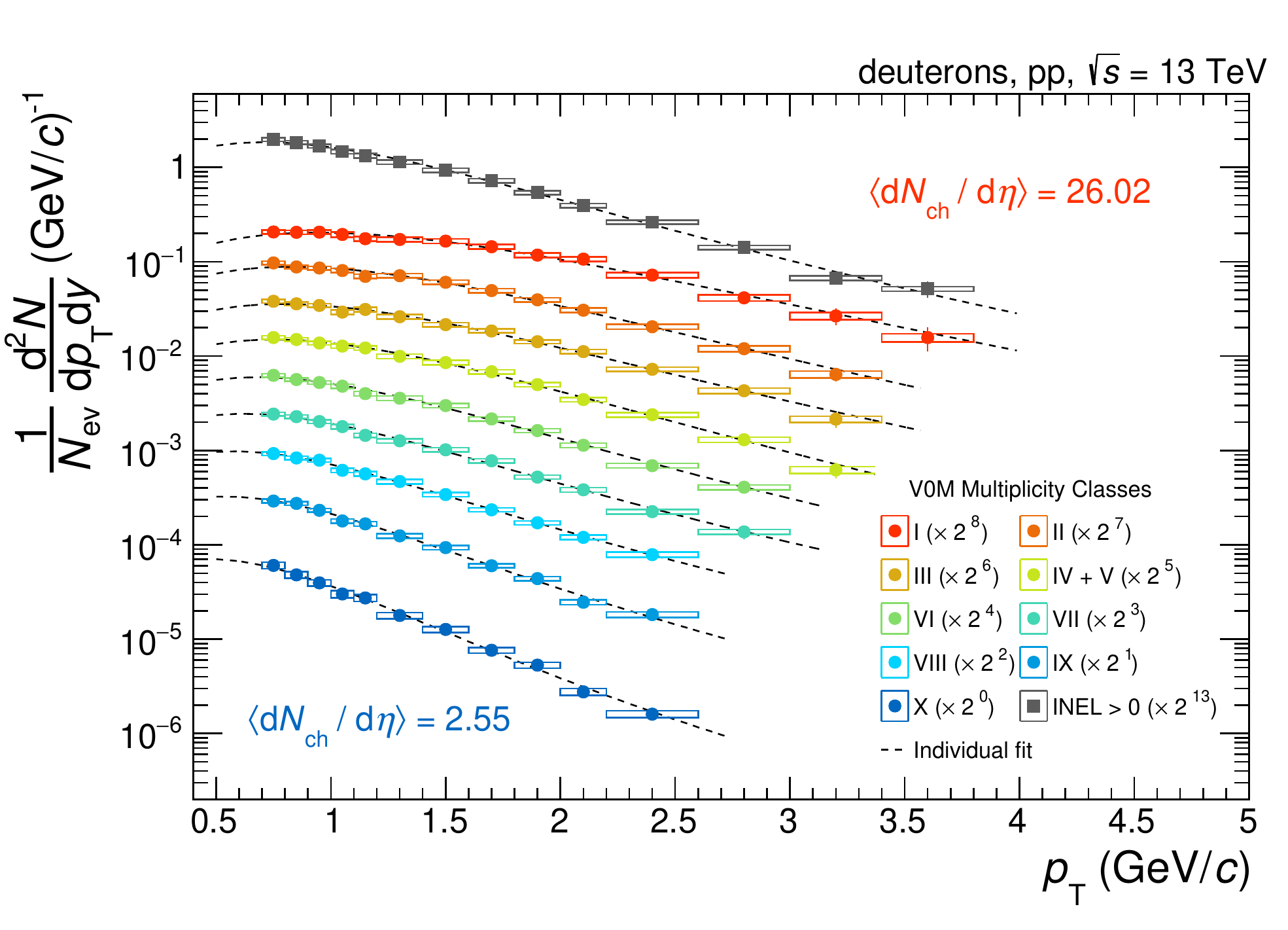}
    \end{subfigure}
    \begin{subfigure}{}
    \centering
        \includegraphics[width=0.7\textwidth]{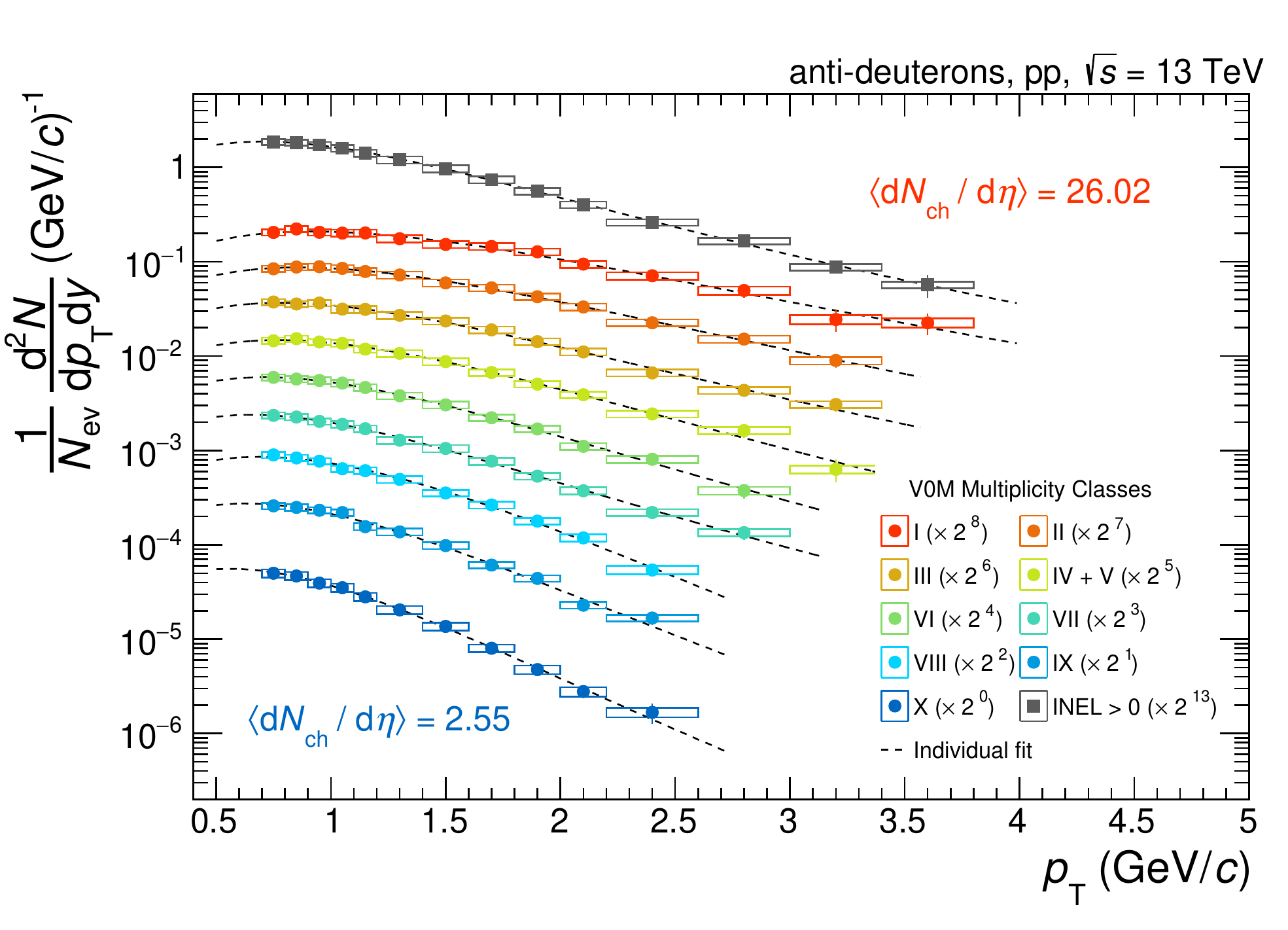}
    \end{subfigure}
     \caption{Transverse-momentum spectra of deuterons (top) and anti-deuterons (bottom) measured in pp collisions at $\sqrt{s}~=~$13~TeV in different multiplicity classes (circles) and in INEL$>$0 events (squares). The mean charged-particle multiplicity for classes I and X are reported in the figures and all the values for the multiplicity classes can be found in Table \ref{tab:yields}. For the analyses in multiplicity classes, the multiplicity increases moving from the bottom of the figure upwards. The statistical uncertainties are represented by vertical bars while the systematic uncertainties are represented by boxes. The dashed lines are individual fits with a L\'evy-Tsallis function~\cite{Tsallis:1987eu}.}
     \label{fig:spectra}
\end{figure}

\section{Results and Discussion}
\label{sec:results}

\begin{figure}[ht]
	\centering
    \includegraphics[width=0.8\textwidth]{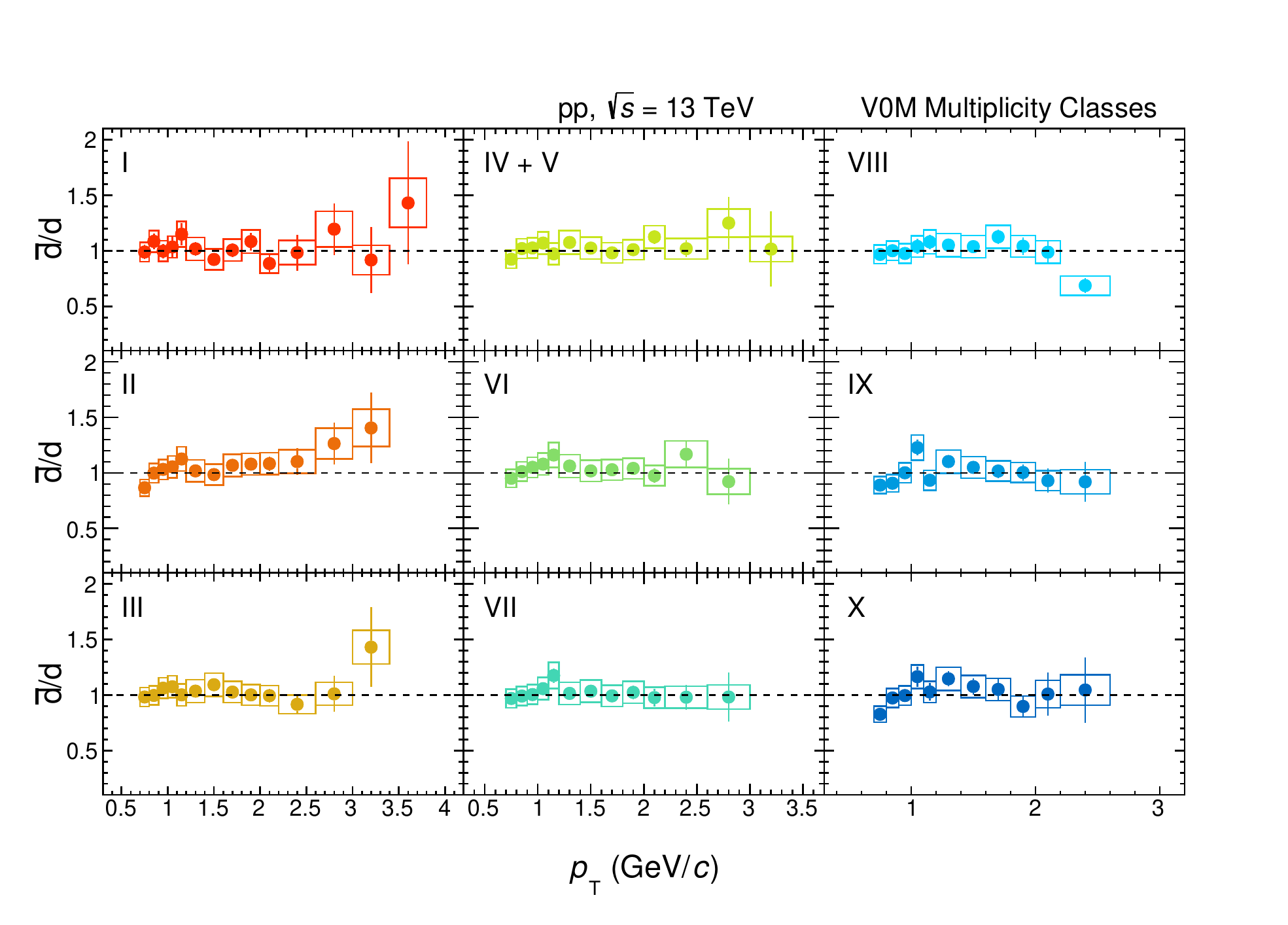}
     \caption{Ratio between the transverse momentum spectra of anti-deuterons and deuterons in different multiplicity classes. The statistical uncertainties are represented by vertical bars while the systematic uncertainties are represented by boxes.}
     \label{fig:ratio}
\end{figure}

The transverse momentum spectra of deuterons and anti-deuterons in different multiplicity classes as well as INEL$>$0 pp collisions are reported in Figure \ref{fig:spectra}. The spectra normalised to inelastic pp collisions (INEL) are included in the data provided with this paper. The mean charged-particle multiplicity $\left<\dndeta\right>$ for each class is reported in Table \ref{tab:yields}. The spectra exhibit a slight hardening with increasing multiplicity: the slope of the spectra becomes less steep and the mean transverse momentum $\left< p_{\mathrm{T}}\right>$ moves towards higher values. This effect is similar to that observed in Pb--Pb collisions, where it is explained with the presence of increasing radial flow with centrality \cite{Adam:2015vda,Acharya:2017dmc}. However, in pp collisions the intensity of the hardening is not as dramatic. The ratio between the spectra of anti-deuterons and deuterons for all the multiplicity classes under study is reported in Figure \ref{fig:ratio}. The ratio is compatible within uncertainties with unity in all multiplicity classes.

To calculate the integrated yield (\dndy) and the mean \pt the spectra have been fitted with the L\'evy-Tsallis function \cite{Tsallis:1987eu,PhysRevLett.84.2770,adams2005k}:
\begin{equation}
	\frac{\mathrm{d}^2N}{\mathrm{d}y \, \mathrm{d}p_{\mathrm{T}}} = \frac{\mathrm{d}N}{\mathrm{d}y}\frac{p_{\mathrm{T}}\left(n-1\right)\left(n-2\right)}{nC [nC + m\left(n-2\right)]}\left(1+\frac{m_{\mathrm{T}}-m}{nC}\right)^{-n},
\end{equation}
where $m$ is the particle rest mass (i.e. the mass of the deuteron), $m_{\mathrm{T}}=\sqrt{m^2+p_{\mathrm{T}}^2}$ is the transverse mass, while $n$, \dndy and $C$ are free fit parameters. The L\'evy-Tsallis function is used to extrapolate the spectra in the unmeasured regions of \pt.
One contribution to the systematic uncertainty is obtained by shifting the data points to the upper border of their systematic uncertainty and to the corresponding lower border. The difference between these values and the reference one is taken as an uncertainty which amounts to $\sim$ 11\%. Another contribution to the systematic uncertainty is estimated by using alternative fit functions such as simple exponentials depending on \pt and \mt, as well as a Boltzmann function, and is found to be $\sim$ 3\%.The two contributions are summed in quadrature. The extrapolation amounts to 25\% of the total yield in the highest multiplicity class, where the widest \pt range is measured, and increases up to 35\% in the lowest multiplicity class.

The statistical uncertainty on the integrated yield is obtained by moving the data points randomly within their statistical uncertainties, using a Gaussian probability distribution centered at the measured data point, with a standard deviation corresponding to the statistical uncertainty. In the unmeasured regions at low and high \pt, the value of the fit function at a given \pt is considered. In this case the statistical uncertainty is estimated using a Monte Carlo method to propagate the uncertainties on the fit parameters.
Following the same procedure, the \meanpt and its statistical and systematic uncertainties are computed.
The resulting mean \pt and \dndy, as well as the parameters of the individual L\'evy-Tsallis fits, are listed in Table \ref{tab:yields}.

\begin{table}
\centering
\caption{\label{tab:yields}Summary of the relevant information about the multiplicity classes and the fits to the measured transverse momentum spectra of anti-deuterons. $\langle \mathrm{d}N_{ch}/\mathrm{d}\eta \rangle$ is the mean pseudorapidity density of the primary charged particles \cite{Acharya:2019kyh}. $n$ and $C$ are the parameters of the L\'evy-Tsallis fit function \cite{Tsallis:1987eu}. $\mathrm{d}N/\mathrm{d}y$ is the integrated yield, with statistical uncertainties, multiplicity-uncorrelated and multiplicity-correlated systematic uncertainties (see the text for details). $\langle p_{\mathrm{T}} \rangle$ is the mean transverse momentum. 
}
\resizebox{\textwidth}{!}{
\begin{tabular}{c|c|c|c|c|c}
\hline
Multiplicity & \multirow{2}{*}{$\langle \mathrm{d}N_{ch}/\mathrm{d}\eta \rangle$} & \multirow{2}{*}{$n$} & \multirow{2}{*}{$C$ (GeV)} & \multirow{2}{*}{$\mathrm{d}N/\mathrm{d}y \left(\times 10^{-4}\right)$} & \multirow{2}{*}{$\langle p_{\mathrm{T}} \rangle$ (GeV/\textit{c})} \\
class        &                                                                    &                      &                            &                                                                        &                                                                    \\ \hline
I            & 26.02 $\pm$ 0.35         & 7  $\pm$ 3     & 0.37 $\pm$ 0.05   & 16.0 $\pm$ 0.4 $\pm$ 0.5 $\pm$ 1.8	& 1.57 $\pm$ 0.08 $\pm$ 0.05 $\pm$ 0.03\\
II           & 20.02 $\pm$ 0.27         & 7  $\pm$ 3     & 0.32 $\pm$ 0.04   & 12.2 $\pm$ 0.2 $\pm$ 0.4 $\pm$ 1.4	& 1.43 $\pm$ 0.04	$\pm$ 0.04 $\pm$ 0.02\\
III          & 16.17 $\pm$ 0.22         & 6  $\pm$ 2     & 0.27 $\pm$ 0.03   & 9.4 $\pm$ 0.1 $\pm$ 0. 3 $\pm$ 1.1	& 1.31 $\pm$ 0.03 $\pm$ 0.03 $\pm$ 0.04\\
IV + V       & 12.91 $\pm$ 0.13         & 8  $\pm$ 3     & 0.27 $\pm$ 0.03   & 7.13 $\pm$ 0.08 $\pm$ 0.20 $\pm$ 0.79	& 1.21 $\pm$ 0.02 $\pm$ 0.01 $\pm$ 0.03\\
VI           & 10.02 $\pm$ 0.14         & 7  $\pm$ 2     & 0.23 $\pm$ 0.03   & 5.34 $\pm$ 0.07 $\pm$ 0.20 $\pm$ 0.59	& 1.12 $\pm$ 0.02 $\pm$ 0.01 $\pm$ 0.03\\
VII          & 7.95  $\pm$ 0.11         & 6  $\pm$ 2     & 0.19 $\pm$ 0.03   & 3.99 $\pm$ 0.07 $\pm$ 0.20 $\pm$ 0.44	& 1.06 $\pm$ 0.02 $\pm$ 0.01 $\pm$ 0.03\\
VIII         & 6.32  $\pm$ 0.09         & 17 $\pm$ 13    & 0.23 $\pm$ 0.03   & 2.73 $\pm$ 0.04 $\pm$ 0.06 $\pm$ 0.30	& 0.98 $\pm$ 0.01 $\pm$ 0.01 $\pm$ 0.03\\
IX           & 4.50  $\pm$ 0.07         & 10 $\pm$ 5     & 0.19 $\pm$ 0.03   &  1.64 $\pm$ 0.03 $\pm$ 0.06 $\pm$ 0.19	& 0.92 $\pm$ 0.01 $\pm$ 0.01 $\pm$ 0.03\\
X            & 2.55  $\pm$ 0.04         & 10 $\pm$ 5     & 0.15 $\pm$ 0.02   & 0.59 $\pm$ 0.02 $\pm$ 0.04 $\pm$ 0.07	& 0.82 $\pm$ 0.01 $\pm$ 0.02 $\pm$ 0.02\\
\hline
\end{tabular}
}
\end{table}

\begin{figure}[h]
	\centering
    \includegraphics[width=0.8\textwidth]{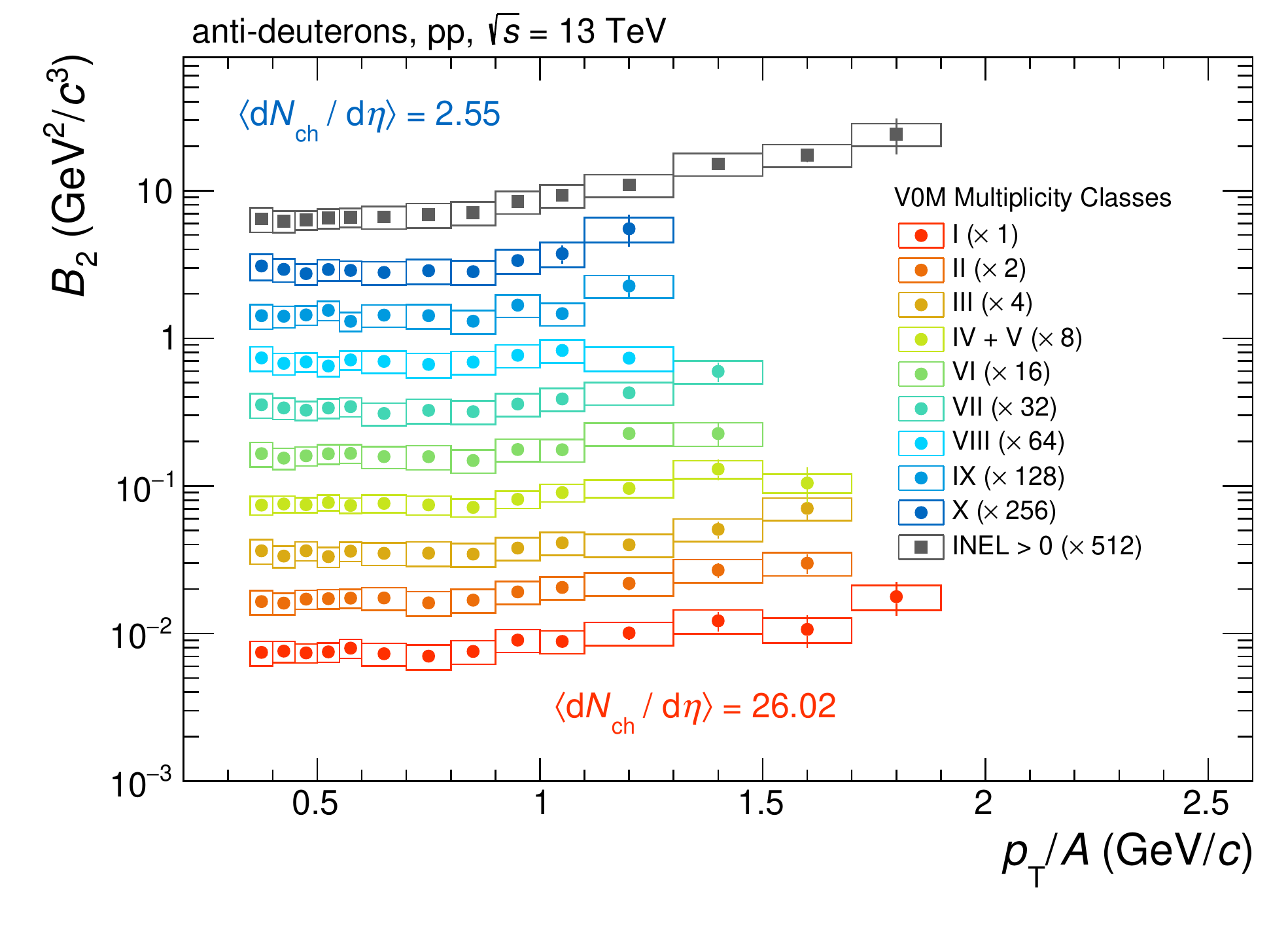}
     \caption{Coalescence parameter $B_2$ for anti-deuterons for different multiplicity classes (circles) and for INEL$>$0 collisions (squares). For the analyses in multiplicity classes, the multiplicity decreases moving from the bottom of the figure upwards. The statistical uncertainties are represented by vertical bars while the systematic uncertainties are represented by boxes. $B_2$ is shown as a function of $\pt/A$, being $A~=~2$ the mass number of the deuteron.}
     \label{fig:b2}
\end{figure}

The coalescence parameter as a function of the transverse momentum is shown in Figure \ref{fig:b2}. The transverse momentum spectra needed for the $B_{2}$ computation are taken from Ref.~\cite{Acharya:2020zji}.
The $B_2$ values for INEL$>$0 collisions show a significant deviation from a transverse momentum independent coalescence parameter as expected by the simplest implementation of the coalescence model. However, it has been shown \cite{Acharya:2019rgc} that the the multiplicity-integrated coalescence parameter is distorted because deuterons are biased more towards higher multiplicity than protons, and consequently have harder \pt spectra than expected from inclusive protons.
The coalescence parameter evaluated in fine multiplicity classes is consistent with a flat behaviour, in agreement with the expectation of the simple coalescence model.

The evolution of the coalescence parameter as a function of the charged particle multiplicity is sensitive to the production mechanism of deuterons.
Recent formulations of the coalescence model \cite{Scheibl:1998tk,Bellini:2018epz} implement an interplay between the size of the collision system and the size of the light nuclei produced via coalescence.

\begin{figure}[h]
	\centering
    \includegraphics[width=0.8\textwidth]{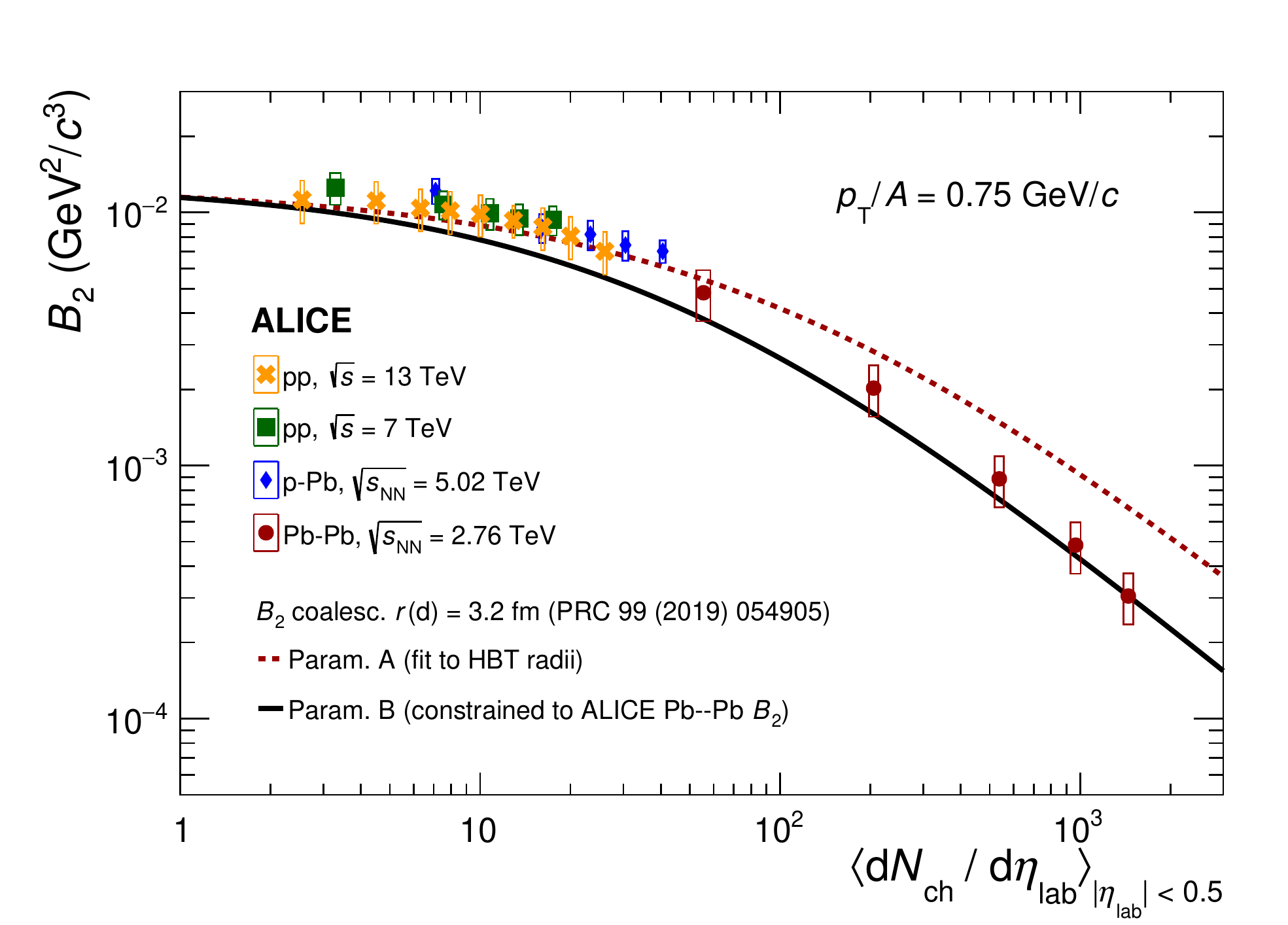}
     \caption{Coalescence parameter $B_2$ at $p_{\mathrm{T}}/A = $ 0.75 GeV/\textit{c} as a function of multiplicity in pp collisions at $\sqrt{s}~=~13$~TeV (anti-deuterons) and in $\sqrt{s}~=~7$~TeV~\cite{Acharya:2019rgc} (average of deuterons and anti-deuterons), in p--Pb collisions at $\snn~=~5.02$~TeV~\cite{Acharya:2019rys} (deuterons) and in Pb--Pb collisions at $\snn~=~2.76$~TeV~\cite{Adam:2015vda} (deuterons). The statistical uncertainties are represented by vertical bars while the systematic uncertainties are represented by boxes. The two lines are theoretical predictions based on two different parameterisations of the HBT radius, see text for details.}
     \label{fig:b2vsmult}
\end{figure}

Figure~\ref{fig:b2vsmult} shows how the $B_2$, for a fixed transverse momentum interval, evolves in different systems as a function of the charged particle multiplicity. $B_2$ is shown at $\pt~=~0.75$~GeV$/c$, which was measured in all the analyses. However, the trend is the same for other \pt values. 
The measurements are compared with the model descriptions detailed in \cite{Bellini:2018epz}.
The two descriptions use different parameterisations for the size of the source. Parameterisation A uses the ALICE measurements of system radii $R$ from HBT studies as a function of multiplicity\cite{abelev2013charged}. These values are fitted with the function:
\begin{equation}
    R = a \; \langle\mathrm{d}N/\mathrm{d}\eta\rangle^{1/3} + b,
    \label{eq:HBT}
\end{equation}
where $a$ and $b$ are free parameters. In Parameterisation B the free parameters $a$ and $b$ in Eq.~\ref{eq:HBT} are fixed to reproduce the $B_2$ of deuterons in \PbPb collisions at $\snn=2.76$ TeV in the centrality class 0--10\%.
The first parameterisation (dashed red line) describes well the measured $B_2$ in pp and p--Pb collisions, while it overestimates the measurements in \PbPb collisions.
However, as outlined by the authors in \cite{Bellini:2018epz}, a more refined parameterisation of the HBT radius evolution through different systems might reduce the observed discrepancy.
The parameterisation of the source size fixed to the $B_2$ measurement in central $\text{\PbPb}$ collisions already departs from the measurements in peripheral \PbPb collisions and it underestimates the coalescence parameter for small colliding systems.

\begin{figure}[h]
	\centering
    \includegraphics[width=0.8\textwidth]{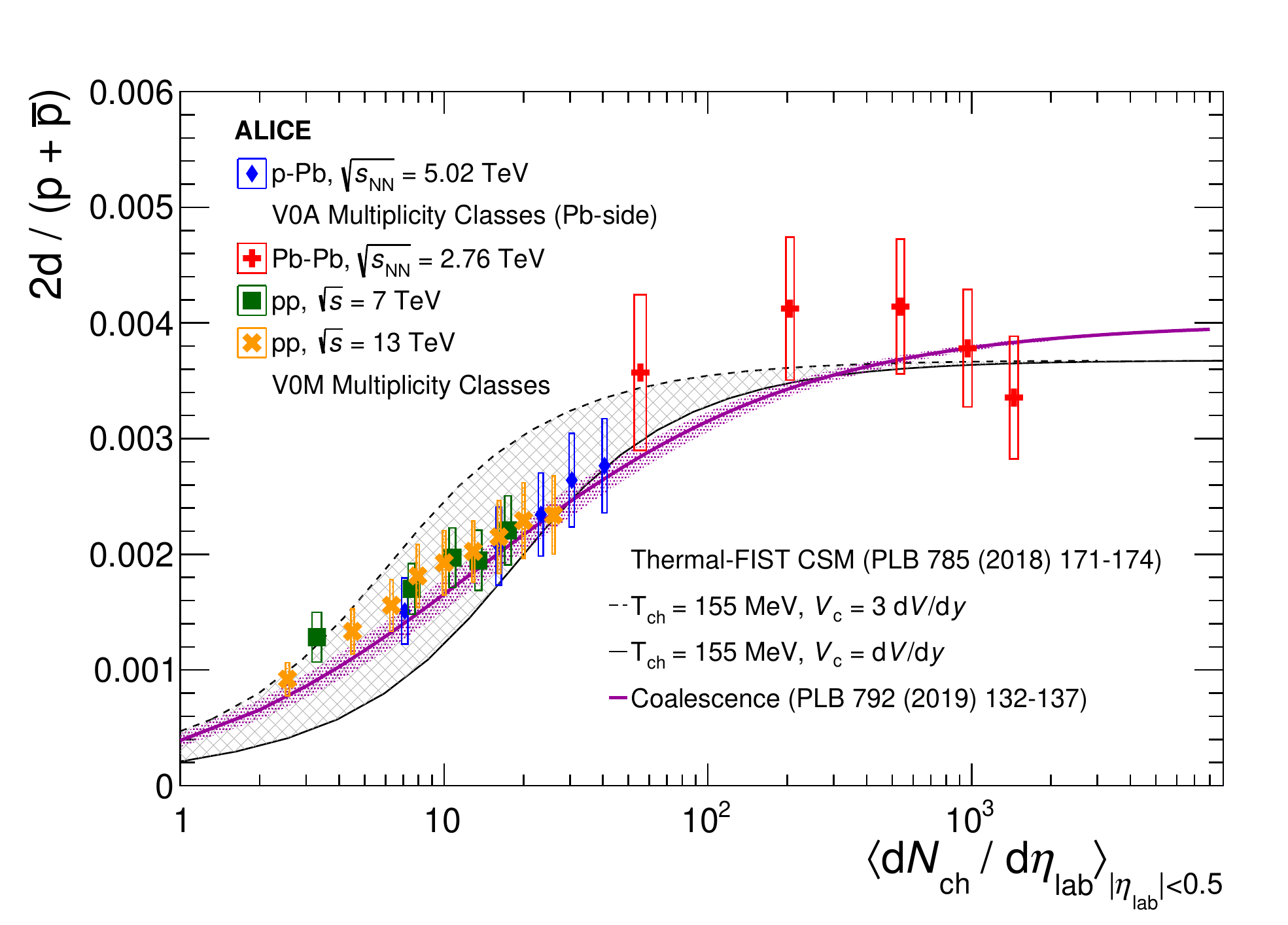}
     \caption{
     Ratio between the \pt-integrated yields of deuterons and protons (sum of protons and anti-protons) for different multiplicities in pp collisions at $\sqrt{s}~=~13$~TeV (anti-deuterons) and in $\sqrt{s}~=~7$~TeV~\cite{Acharya:2019rgc} (deuterons), in p--Pb collisions at $\snn~=~5.02$~TeV~\cite{Acharya:2019rys} (deuterons) and in Pb--Pb collisions at $\snn~=~2.76$~TeV~\cite{Adam:2015vda} (deuterons). The statistical uncertainties are represented by vertical bars while the systematic uncertainties are represented by boxes. The two black lines are the theoretical predictions of the Thermal-FIST statistical model \cite{Vovchenko:2018fiy} for two sizes of the correlation volume $V_C$, while the magenta line represents the expectation from a coalescence model \cite{Sun:2018mqq}.
     }
     \label{fig:DoP}
\end{figure}

Figure~\ref{fig:DoP} shows the ratio of the \pt-integrated yields of deuterons and protons for different multiplicities in different collisions systems and at different energies.
The ratio increases monotonically with multiplicity for pp and p--Pb collisions and eventually saturates for Pb--Pb collisions. The experimental data are compared with a SHM prediction. 
In this implementation of the model, called the Canonical Statistical Model (CSM), exact conservation of baryon number ($B$), charge ($Q$), and strangeness ($S$) is enforced using the recently developed THERMAL-FIST package~\cite{Vovchenko:2018fiy}. The calculations with the CSM are performed using 155~MeV for the chemical freeze-out temperature, $B$ = $Q$ = $S$ = 0 and two different values of the correlation volume, which is expressed in terms of rapidity units \dvdy, corresponding to one and three units of rapidity, respectively.
The model qualitatively reproduces the trend observed in data. This might suggest that for small collision systems the light (anti-) nuclei production could be canonically suppressed and that a canonical correlation volume might exist. The correlation volume required to describe the measurements is larger than one unit of rapidity.
However, such a canonical suppression should also affect the p/$\pi$ ratio in a similar way and this is not observed in the experimental measurements \cite{Abelev:2013vea,Acharya:2018orn}.

A full coalescence calculation, taking into account the interplay between the system size and the width of the wave function of the produced $\text{(anti-)deuterons}$, is also able to describe the measured trend of the d/p ratio~\cite{Sun:2018mqq} and it describes the data consistently better than CSM for all system sizes.

\section{Conclusions}
\label{sec:conclusions}
The results on (anti-)deuteron production presented in this paper display a smooth evolution with multiplicity across different reaction systems, in agreement with the measurements of other light-flavoured hadrons. This suggests that a common physics process might be able to describe the nuclei production in all hadronic collision systems.
Coalescence and statistical hadronisation models are able to describe qualitatively the observed trend in the d/p ratio and $B_2$ as a function of the charged particle multiplicity. However, with the precision of the current measurements it is not possible to distinguish which mechanism drives the (anti-)deuteron production. On the other hand, it is not clear whether the CSM would be able to describe simultaneously the d/p and the p/$\pi$ ratios with the same chemical freeze-out conditions.

No substantial differences are seen in the dependence of nuclei production on the charged multiplicity in pp and p--Pb collisions and with the Pb--Pb data sample collected in Run 2 it will be also possible to perform a direct comparison with peripheral Pb--Pb collisions.
With the enhanced luminosity in Run 3, it will be possible to measure pp collisions with multiplicities similar to those observed in mid-central \mbox{Pb--Pb} collisions. It will be interesting to see whether ALICE can confirm this dependence when measuring nuclei production in pp and Pb--Pb collisions at the same multiplicity.
%
%
%
%

\newenvironment{acknowledgement}{\relax}{\relax}
\begin{acknowledgement}
\section*{Acknowledgements}

The ALICE Collaboration would like to thank all its engineers and technicians for their invaluable contributions to the construction of the experiment and the CERN accelerator teams for the outstanding performance of the LHC complex.
The ALICE Collaboration gratefully acknowledges the resources and support provided by all Grid centres and the Worldwide LHC Computing Grid (WLCG) collaboration.
The ALICE Collaboration acknowledges the following funding agencies for their support in building and running the ALICE detector:
A. I. Alikhanyan National Science Laboratory (Yerevan Physics Institute) Foundation (ANSL), State Committee of Science and World Federation of Scientists (WFS), Armenia;
Austrian Academy of Sciences, Austrian Science Fund (FWF): [M 2467-N36] and Nationalstiftung f\"{u}r Forschung, Technologie und Entwicklung, Austria;
Ministry of Communications and High Technologies, National Nuclear Research Center, Azerbaijan;
Conselho Nacional de Desenvolvimento Cient\'{\i}fico e Tecnol\'{o}gico (CNPq), Financiadora de Estudos e Projetos (Finep), Funda\c{c}\~{a}o de Amparo \`{a} Pesquisa do Estado de S\~{a}o Paulo (FAPESP) and Universidade Federal do Rio Grande do Sul (UFRGS), Brazil;
Ministry of Education of China (MOEC) , Ministry of Science \& Technology of China (MSTC) and National Natural Science Foundation of China (NSFC), China;
Ministry of Science and Education and Croatian Science Foundation, Croatia;
Centro de Aplicaciones Tecnol\'{o}gicas y Desarrollo Nuclear (CEADEN), Cubaenerg\'{\i}a, Cuba;
Ministry of Education, Youth and Sports of the Czech Republic, Czech Republic;
The Danish Council for Independent Research | Natural Sciences, the VILLUM FONDEN and Danish National Research Foundation (DNRF), Denmark;
Helsinki Institute of Physics (HIP), Finland;
Commissariat \`{a} l'Energie Atomique (CEA), Institut National de Physique Nucl\'{e}aire et de Physique des Particules (IN2P3) and Centre National de la Recherche Scientifique (CNRS) and R\'{e}gion des  Pays de la Loire, France;
Bundesministerium f\"{u}r Bildung und Forschung (BMBF) and GSI Helmholtzzentrum f\"{u}r Schwerionenforschung GmbH, Germany;
General Secretariat for Research and Technology, Ministry of Education, Research and Religions, Greece;
National Research, Development and Innovation Office, Hungary;
Department of Atomic Energy Government of India (DAE), Department of Science and Technology, Government of India (DST), University Grants Commission, Government of India (UGC) and Council of Scientific and Industrial Research (CSIR), India;
Indonesian Institute of Science, Indonesia;
Centro Fermi - Museo Storico della Fisica e Centro Studi e Ricerche Enrico Fermi and Istituto Nazionale di Fisica Nucleare (INFN), Italy;
Institute for Innovative Science and Technology , Nagasaki Institute of Applied Science (IIST), Japanese Ministry of Education, Culture, Sports, Science and Technology (MEXT) and Japan Society for the Promotion of Science (JSPS) KAKENHI, Japan;
Consejo Nacional de Ciencia (CONACYT) y Tecnolog\'{i}a, through Fondo de Cooperaci\'{o}n Internacional en Ciencia y Tecnolog\'{i}a (FONCICYT) and Direcci\'{o}n General de Asuntos del Personal Academico (DGAPA), Mexico;
Nederlandse Organisatie voor Wetenschappelijk Onderzoek (NWO), Netherlands;
The Research Council of Norway, Norway;
Commission on Science and Technology for Sustainable Development in the South (COMSATS), Pakistan;
Pontificia Universidad Cat\'{o}lica del Per\'{u}, Peru;
Ministry of Science and Higher Education and National Science Centre, Poland;
Korea Institute of Science and Technology Information and National Research Foundation of Korea (NRF), Republic of Korea;
Ministry of Education and Scientific Research, Institute of Atomic Physics and Ministry of Research and Innovation and Institute of Atomic Physics, Romania;
Joint Institute for Nuclear Research (JINR), Ministry of Education and Science of the Russian Federation, National Research Centre Kurchatov Institute, Russian Science Foundation and Russian Foundation for Basic Research, Russia;
Ministry of Education, Science, Research and Sport of the Slovak Republic, Slovakia;
National Research Foundation of South Africa, South Africa;
Swedish Research Council (VR) and Knut \& Alice Wallenberg Foundation (KAW), Sweden;
European Organization for Nuclear Research, Switzerland;
Suranaree University of Technology (SUT), National Science and Technology Development Agency (NSDTA) and Office of the Higher Education Commission under NRU project of Thailand, Thailand;
Turkish Atomic Energy Agency (TAEK), Turkey;
National Academy of  Sciences of Ukraine, Ukraine;
Science and Technology Facilities Council (STFC), United Kingdom;
National Science Foundation of the United States of America (NSF) and United States Department of Energy, Office of Nuclear Physics (DOE NP), United States of America.    
\end{acknowledgement}

\bibliographystyle{utphys}   
\bibliography{biblio}

\newpage
\appendix
\section{The ALICE Collaboration}
\label{app:collab}

\begingroup
\small
\begin{flushleft}
S.~Acharya\Irefn{org141}\And 
D.~Adamov\'{a}\Irefn{org94}\And 
A.~Adler\Irefn{org74}\And 
J.~Adolfsson\Irefn{org80}\And 
M.M.~Aggarwal\Irefn{org99}\And 
G.~Aglieri Rinella\Irefn{org33}\And 
M.~Agnello\Irefn{org30}\And 
N.~Agrawal\Irefn{org10}\textsuperscript{,}\Irefn{org53}\And 
Z.~Ahammed\Irefn{org141}\And 
S.~Ahmad\Irefn{org16}
% ...
\And
\renewcommand\labelenumi{\textsuperscript{\theenumi}~}

\section*{Affiliation notes}
\renewcommand\theenumi{\roman{enumi}}
\begin{Authlist}
\item \Adef{org*}Deceased
\item \Adef{orgI}Dipartimento DET del Politecnico di Torino, Turin, Italy
\item \Adef{orgII}M.V. Lomonosov Moscow State University, D.V. Skobeltsyn Institute of Nuclear, Physics, Moscow, Russia
\item \Adef{orgIII}Department of Applied Physics, Aligarh Muslim University, Aligarh, India
\item \Adef{orgIV}Institute of Theoretical Physics, University of Wroclaw, Poland
\end{Authlist}

\section*{Collaboration Institutes}
\renewcommand\theenumi{\arabic{enumi}~}
\begin{Authlist}
\item \Idef{org1}A.I. Alikhanyan National Science Laboratory (Yerevan Physics Institute) Foundation, Yerevan, Armenia
\item \Idef{org2}Bogolyubov Institute for Theoretical Physics, National Academy of Sciences of Ukraine, Kiev, Ukraine
\item \Idef{org3}Bose Institute, Department of Physics  and Centre for Astroparticle Physics and Space Science (CAPSS), Kolkata, India
\item \Idef{org4}Budker Institute for Nuclear Physics, Novosibirsk, Russia
\item \Idef{org5}California Polytechnic State University, San Luis Obispo, California, United States
\item \Idef{org6}Central China Normal University, Wuhan, China
\item \Idef{org7}Centre de Calcul de l'IN2P3, Villeurbanne, Lyon, France
\item \Idef{org8}Centro de Aplicaciones Tecnol\'{o}gicas y Desarrollo Nuclear (CEADEN), Havana, Cuba
\item \Idef{org9}Centro de Investigaci\'{o}n y de Estudios Avanzados (CINVESTAV), Mexico City and M\'{e}rida, Mexico
\item \Idef{org10}Centro Fermi - Museo Storico della Fisica e Centro Studi e Ricerche ``Enrico Fermi', Rome, Italy
\item \Idef{org11}Chicago State University, Chicago, Illinois, United States
\item \Idef{org12}China Institute of Atomic Energy, Beijing, China
\item \Idef{org13}Comenius University Bratislava, Faculty of Mathematics, Physics and Informatics, Bratislava, Slovakia
\item \Idef{org14}COMSATS University Islamabad, Islamabad, Pakistan
\item \Idef{org15}Creighton University, Omaha, Nebraska, United States
\item \Idef{org16}Department of Physics, Aligarh Muslim University, Aligarh, India
\item \Idef{org17}Department of Physics, Pusan National University, Pusan, Republic of Korea
\item \Idef{org18}Department of Physics, Sejong University, Seoul, Republic of Korea
\item \Idef{org19}Department of Physics, University of California, Berkeley, California, United States
\item \Idef{org20}Department of Physics, University of Oslo, Oslo, Norway
\item \Idef{org21}Department of Physics and Technology, University of Bergen, Bergen, Norway
\item \Idef{org22}Dipartimento di Fisica dell'Universit\`{a} 'La Sapienza' and Sezione INFN, Rome, Italy
\item \Idef{org23}Dipartimento di Fisica dell'Universit\`{a} and Sezione INFN, Cagliari, Italy
\item \Idef{org24}Dipartimento di Fisica dell'Universit\`{a} and Sezione INFN, Trieste, Italy
\item \Idef{org25}Dipartimento di Fisica dell'Universit\`{a} and Sezione INFN, Turin, Italy
\item \Idef{org26}Dipartimento di Fisica e Astronomia dell'Universit\`{a} and Sezione INFN, Bologna, Italy
\item \Idef{org27}Dipartimento di Fisica e Astronomia dell'Universit\`{a} and Sezione INFN, Catania, Italy
\item \Idef{org28}Dipartimento di Fisica e Astronomia dell'Universit\`{a} and Sezione INFN, Padova, Italy
\item \Idef{org29}Dipartimento di Fisica `E.R.~Caianiello' dell'Universit\`{a} and Gruppo Collegato INFN, Salerno, Italy
\item \Idef{org30}Dipartimento DISAT del Politecnico and Sezione INFN, Turin, Italy
\item \Idef{org31}Dipartimento di Scienze e Innovazione Tecnologica dell'Universit\`{a} del Piemonte Orientale and INFN Sezione di Torino, Alessandria, Italy
\item \Idef{org32}Dipartimento Interateneo di Fisica `M.~Merlin' and Sezione INFN, Bari, Italy
\item \Idef{org33}European Organization for Nuclear Research (CERN), Geneva, Switzerland
\item \Idef{org34}Faculty of Electrical Engineering, Mechanical Engineering and Naval Architecture, University of Split, Split, Croatia
\item \Idef{org35}Faculty of Engineering and Science, Western Norway University of Applied Sciences, Bergen, Norway
\item \Idef{org36}Faculty of Nuclear Sciences and Physical Engineering, Czech Technical University in Prague, Prague, Czech Republic
\item \Idef{org37}Faculty of Science, P.J.~\v{S}af\'{a}rik University, Ko\v{s}ice, Slovakia
\item \Idef{org38}Frankfurt Institute for Advanced Studies, Johann Wolfgang Goethe-Universit\"{a}t Frankfurt, Frankfurt, Germany
\item \Idef{org39}Fudan University, Shanghai, China
\item \Idef{org40}Gangneung-Wonju National University, Gangneung, Republic of Korea
\item \Idef{org41}Gauhati University, Department of Physics, Guwahati, India
\item \Idef{org42}Helmholtz-Institut f\"{u}r Strahlen- und Kernphysik, Rheinische Friedrich-Wilhelms-Universit\"{a}t Bonn, Bonn, Germany
\item \Idef{org43}Helsinki Institute of Physics (HIP), Helsinki, Finland
\item \Idef{org44}High Energy Physics Group,  Universidad Aut\'{o}noma de Puebla, Puebla, Mexico
\item \Idef{org45}Hiroshima University, Hiroshima, Japan
\item \Idef{org46}Hochschule Worms, Zentrum  f\"{u}r Technologietransfer und Telekommunikation (ZTT), Worms, Germany
\item \Idef{org47}Horia Hulubei National Institute of Physics and Nuclear Engineering, Bucharest, Romania
\item \Idef{org48}Indian Institute of Technology Bombay (IIT), Mumbai, India
\item \Idef{org49}Indian Institute of Technology Indore, Indore, India
\item \Idef{org50}Indonesian Institute of Sciences, Jakarta, Indonesia
\item \Idef{org51}INFN, Laboratori Nazionali di Frascati, Frascati, Italy
\item \Idef{org52}INFN, Sezione di Bari, Bari, Italy
\item \Idef{org53}INFN, Sezione di Bologna, Bologna, Italy
\item \Idef{org54}INFN, Sezione di Cagliari, Cagliari, Italy
\item \Idef{org55}INFN, Sezione di Catania, Catania, Italy
\item \Idef{org56}INFN, Sezione di Padova, Padova, Italy
\item \Idef{org57}INFN, Sezione di Roma, Rome, Italy
\item \Idef{org58}INFN, Sezione di Torino, Turin, Italy
\item \Idef{org59}INFN, Sezione di Trieste, Trieste, Italy
\item \Idef{org60}Inha University, Incheon, Republic of Korea
\item \Idef{org61}Institut de Physique Nucl\'{e}aire d'Orsay (IPNO), Institut National de Physique Nucl\'{e}aire et de Physique des Particules (IN2P3/CNRS), Universit\'{e} de Paris-Sud, Universit\'{e} Paris-Saclay, Orsay, France
\item \Idef{org62}Institute for Nuclear Research, Academy of Sciences, Moscow, Russia
\item \Idef{org63}Institute for Subatomic Physics, Utrecht University/Nikhef, Utrecht, Netherlands
\item \Idef{org64}Institute of Experimental Physics, Slovak Academy of Sciences, Ko\v{s}ice, Slovakia
\item \Idef{org65}Institute of Physics, Homi Bhabha National Institute, Bhubaneswar, India
\item \Idef{org66}Institute of Physics of the Czech Academy of Sciences, Prague, Czech Republic
\item \Idef{org67}Institute of Space Science (ISS), Bucharest, Romania
\item \Idef{org68}Institut f\"{u}r Kernphysik, Johann Wolfgang Goethe-Universit\"{a}t Frankfurt, Frankfurt, Germany
\item \Idef{org69}Instituto de Ciencias Nucleares, Universidad Nacional Aut\'{o}noma de M\'{e}xico, Mexico City, Mexico
\item \Idef{org70}Instituto de F\'{i}sica, Universidade Federal do Rio Grande do Sul (UFRGS), Porto Alegre, Brazil
\item \Idef{org71}Instituto de F\'{\i}sica, Universidad Nacional Aut\'{o}noma de M\'{e}xico, Mexico City, Mexico
\item \Idef{org72}iThemba LABS, National Research Foundation, Somerset West, South Africa
\item \Idef{org73}Jeonbuk National University, Jeonju, Republic of Korea
\item \Idef{org74}Johann-Wolfgang-Goethe Universit\"{a}t Frankfurt Institut f\"{u}r Informatik, Fachbereich Informatik und Mathematik, Frankfurt, Germany
\item \Idef{org75}Joint Institute for Nuclear Research (JINR), Dubna, Russia
\item \Idef{org76}Korea Institute of Science and Technology Information, Daejeon, Republic of Korea
\item \Idef{org77}KTO Karatay University, Konya, Turkey
\item \Idef{org78}Laboratoire de Physique Subatomique et de Cosmologie, Universit\'{e} Grenoble-Alpes, CNRS-IN2P3, Grenoble, France
\item \Idef{org79}Lawrence Berkeley National Laboratory, Berkeley, California, United States
\item \Idef{org80}Lund University Department of Physics, Division of Particle Physics, Lund, Sweden
\item \Idef{org81}Nagasaki Institute of Applied Science, Nagasaki, Japan
\item \Idef{org82}Nara Women{'}s University (NWU), Nara, Japan
\item \Idef{org83}National and Kapodistrian University of Athens, School of Science, Department of Physics , Athens, Greece
\item \Idef{org84}National Centre for Nuclear Research, Warsaw, Poland
\item \Idef{org85}National Institute of Science Education and Research, Homi Bhabha National Institute, Jatni, India
\item \Idef{org86}National Nuclear Research Center, Baku, Azerbaijan
\item \Idef{org87}National Research Centre Kurchatov Institute, Moscow, Russia
\item \Idef{org88}Niels Bohr Institute, University of Copenhagen, Copenhagen, Denmark
\item \Idef{org89}Nikhef, National institute for subatomic physics, Amsterdam, Netherlands
\item \Idef{org90}NRC Kurchatov Institute IHEP, Protvino, Russia
\item \Idef{org91}NRC «Kurchatov Institute»  - ITEP, Moscow, Russia
\item \Idef{org92}NRNU Moscow Engineering Physics Institute, Moscow, Russia
\item \Idef{org93}Nuclear Physics Group, STFC Daresbury Laboratory, Daresbury, United Kingdom
\item \Idef{org94}Nuclear Physics Institute of the Czech Academy of Sciences, \v{R}e\v{z} u Prahy, Czech Republic
\item \Idef{org95}Oak Ridge National Laboratory, Oak Ridge, Tennessee, United States
\item \Idef{org96}Ohio State University, Columbus, Ohio, United States
\item \Idef{org97}Petersburg Nuclear Physics Institute, Gatchina, Russia
\item \Idef{org98}Physics department, Faculty of science, University of Zagreb, Zagreb, Croatia
\item \Idef{org99}Physics Department, Panjab University, Chandigarh, India
\item \Idef{org100}Physics Department, University of Jammu, Jammu, India
\item \Idef{org101}Physics Department, University of Rajasthan, Jaipur, India
\item \Idef{org102}Physikalisches Institut, Eberhard-Karls-Universit\"{a}t T\"{u}bingen, T\"{u}bingen, Germany
\item \Idef{org103}Physikalisches Institut, Ruprecht-Karls-Universit\"{a}t Heidelberg, Heidelberg, Germany
\item \Idef{org104}Physik Department, Technische Universit\"{a}t M\"{u}nchen, Munich, Germany
\item \Idef{org105}Politecnico di Bari, Bari, Italy
\item \Idef{org106}Research Division and ExtreMe Matter Institute EMMI, GSI Helmholtzzentrum f\"ur Schwerionenforschung GmbH, Darmstadt, Germany
\item \Idef{org107}Rudjer Bo\v{s}kovi\'{c} Institute, Zagreb, Croatia
\item \Idef{org108}Russian Federal Nuclear Center (VNIIEF), Sarov, Russia
\item \Idef{org109}Saha Institute of Nuclear Physics, Homi Bhabha National Institute, Kolkata, India
\item \Idef{org110}School of Physics and Astronomy, University of Birmingham, Birmingham, United Kingdom
\item \Idef{org111}Secci\'{o}n F\'{\i}sica, Departamento de Ciencias, Pontificia Universidad Cat\'{o}lica del Per\'{u}, Lima, Peru
\item \Idef{org112}St. Petersburg State University, St. Petersburg, Russia
\item \Idef{org113}Stefan Meyer Institut f\"{u}r Subatomare Physik (SMI), Vienna, Austria
\item \Idef{org114}SUBATECH, IMT Atlantique, Universit\'{e} de Nantes, CNRS-IN2P3, Nantes, France
\item \Idef{org115}Suranaree University of Technology, Nakhon Ratchasima, Thailand
\item \Idef{org116}Technical University of Ko\v{s}ice, Ko\v{s}ice, Slovakia
\item \Idef{org117}Technische Universit\"{a}t M\"{u}nchen, Excellence Cluster 'Universe', Munich, Germany
\item \Idef{org118}The Henryk Niewodniczanski Institute of Nuclear Physics, Polish Academy of Sciences, Cracow, Poland
\item \Idef{org119}The University of Texas at Austin, Austin, Texas, United States
\item \Idef{org120}Universidad Aut\'{o}noma de Sinaloa, Culiac\'{a}n, Mexico
\item \Idef{org121}Universidade de S\~{a}o Paulo (USP), S\~{a}o Paulo, Brazil
\item \Idef{org122}Universidade Estadual de Campinas (UNICAMP), Campinas, Brazil
\item \Idef{org123}Universidade Federal do ABC, Santo Andre, Brazil
\item \Idef{org124}University of Cape Town, Cape Town, South Africa
\item \Idef{org125}University of Houston, Houston, Texas, United States
\item \Idef{org126}University of Jyv\"{a}skyl\"{a}, Jyv\"{a}skyl\"{a}, Finland
\item \Idef{org127}University of Liverpool, Liverpool, United Kingdom
\item \Idef{org128}University of Science and Techonology of China, Hefei, China
\item \Idef{org129}University of South-Eastern Norway, Tonsberg, Norway
\item \Idef{org130}University of Tennessee, Knoxville, Tennessee, United States
\item \Idef{org131}University of the Witwatersrand, Johannesburg, South Africa
\item \Idef{org132}University of Tokyo, Tokyo, Japan
\item \Idef{org133}University of Tsukuba, Tsukuba, Japan
\item \Idef{org134}Universit\'{e} Clermont Auvergne, CNRS/IN2P3, LPC, Clermont-Ferrand, France
\item \Idef{org135}Universit\'{e} de Lyon, Universit\'{e} Lyon 1, CNRS/IN2P3, IPN-Lyon, Villeurbanne, Lyon, France
\item \Idef{org136}Universit\'{e} de Strasbourg, CNRS, IPHC UMR 7178, F-67000 Strasbourg, France, Strasbourg, France
\item \Idef{org137}Universit\'{e} Paris-Saclay Centre d'Etudes de Saclay (CEA), IRFU, D\'{e}partment de Physique Nucl\'{e}aire (DPhN), Saclay, France
\item \Idef{org138}Universit\`{a} degli Studi di Foggia, Foggia, Italy
\item \Idef{org139}Universit\`{a} degli Studi di Pavia, Pavia, Italy
\item \Idef{org140}Universit\`{a} di Brescia, Brescia, Italy
\item \Idef{org141}Variable Energy Cyclotron Centre, Homi Bhabha National Institute, Kolkata, India
\item \Idef{org142}Warsaw University of Technology, Warsaw, Poland
\item \Idef{org143}Wayne State University, Detroit, Michigan, United States
\item \Idef{org144}Westf\"{a}lische Wilhelms-Universit\"{a}t M\"{u}nster, Institut f\"{u}r Kernphysik, M\"{u}nster, Germany
\item \Idef{org145}Wigner Research Centre for Physics, Budapest, Hungary
\item \Idef{org146}Yale University, New Haven, Connecticut, United States
\item \Idef{org147}Yonsei University, Seoul, Republic of Korea
\end{Authlist}
\endgroup
\end{document}